\documentclass[nofootinbib,prd,12pt,superscriptaddress]{revtex4}%
\usepackage{amsmath}
\usepackage{amsfonts}
\usepackage{amssymb}
\usepackage{graphicx}%
\usepackage{amsmath,amssymb}
\usepackage{mathrsfs}
\usepackage{graphicx}
\usepackage{color}
\usepackage{subfigure}
\usepackage{fancyhdr}
\usepackage{multirow}
\usepackage{float}
\usepackage{epsfig}
\usepackage{amsfonts}
\usepackage{bm}
\def\be{\begin{equation}}
\def\ee{\end{equation}}
\def\beq{\begin{eqnarray}}
\def\eeq{\end{eqnarray}}

\begin{document}

\title{Charged Wormhole Solutions in 4D Einstein-Gauss-Bonnet Gravity}

\author{Piyachat Panyasiripan}
\email{piyachat.pa@mail.wu.ac.th}
\affiliation{School of Science, Walailak University, Nakhon Si Thammarat, 80160, Thailand}

\author{Fereshteh Felegary}
\email{fereshteh.felegary@gmail.com}
\affiliation{School of Science, Walailak University, Nakhon Si Thammarat, 80160, Thailand}
\affiliation{Faculty of Physics, Shahrood University of Technology, Shahrood, Iran}

\author{Phongpichit Channuie}
\email{phongpichit.ch@mail.wu.ac.th}
\affiliation{School of Science, Walailak University, Nakhon Si Thammarat, 80160, Thailand}
\affiliation{College of Graduate Studies, Walailak University, Nakhon Si Thammarat, 80160, Thailand}

\date{\today}

\begin{abstract}

In the present work, we construct models of static wormholes in 4-dimensional Einstein-Gauss-Bonnet (4D EGB) gravity with an (an)isotropic energy momentum tensor (EMT) and a Maxwell field as supporting matters for the wormhole geometry assuming a constant redshift function. We obtain exact spherically symmetric wormhole solutions in 4D EGB gravity for an isotropic and anisotropic matter sources under the effect of electric charge. The solutions are more generalized in the sense of incorporating the charge contributions. Furthermore, we examine the null, weak and strong energy conditions at the wormhole throat of radius $r=r_{0}$. We demonstrate that at the wormhole throat, the classical energy conditions are violated by arbitrary small amount. Additionally, we analyze the wormhole geometry incorporating semiclassical corrections through embedding diagrams. We discover that the wormhole solutions are in equilibrium as the anisotropic and hydrostatic forces cancel each other.

\end{abstract}

\maketitle


\section{Introduction}
The concept of wormholes has long fascinated scientists and the public alike, offering tantalizing possibilities of shortcuts through spacetime, allowing for rapid travel between distant regions of the universe. In the realm of theoretical physics, wormholes emerge as solutions to Einstein's field equations of general relativity, representing bridges that connect separate points in spacetime. The first theoretical exploration of wormholes attracted considerable attention soon after the discovery of General Relativity, with initial work by Flamm \cite{Flamm:1916}. This investigation was further advanced by Einstein and Rosen \cite{Einstein:1935tc}, who introduced the Einstein-Rosen bridge to describe wormholes at that time. However, the early investigations into wormholes were primarily theoretical exercises with no practical connection to physical reality. This changed when Morris and Thorne conducted their study, demonstrating the potential existence of wormholes and how they could be utilized for interstellar travel \cite{Morris:1988cz}.

Subsequently, inspired by the ideas of Morris and Thorne \cite{Morris:1988cz}, there has been intense activity in investigating wormholes within modified theories of gravity and higher-dimensional gravity theories \cite{Harko:2013yb,Zangeneh:2014noa,Banerjee:2020uyi,Shaikh:2016dpl,Bronnikov:2015pha}. Moreover, there has been research on wormholes in extended theories of gravity \cite{DeFalco:2023twb,Rosa:2018jwp,Rosa:2022osy,Rosa:2023guo,Rosa:2021yym,Rosa:2023olc}, focusing on developing astrophysical techniques to detect them \cite{DeFalco:2020afv,DeFalco:2021klh,DeFalco:2021btn,DeFalco:2023kqy}, the reconstruction of their solutions \cite{DeFalco:2021ksd}, and their existence in higher dimensions \cite{Deshpande:2022zfm}. Among many other interesting scenarios, wormhole solutions in Einstein-Gauss-Bonnet theory represent a fascinating extension of the traditional wormhole concept by incorporating higher-dimensional and higher-order curvature terms, see Refs.\cite{Bhawal:1992sz,Mehdizadeh:2020nrw,Ilyas:2023rde,Zubair:2023wjh,Rani:2016xpa,Mehdizadeh:2015jra,Dotti:2007az,Maeda:2008nz}. However, in 4D, the Gauss-Bonnet (GB) term is a topological invariant and does not affect gravitational dynamics. Recently, Glavan and Lin \cite{Glavan:2019inb} proposed a method to regularize the Gauss-Bonnet equations dimensionally, resulting in a 4D metric theory that circumvents Lovelock's theorem and avoids Ostrogradsky instability. Their approach involves formulating the equations in D dimensions, rescaling the coupling constant 
$\alpha \to \alpha/(D -4)$, and then taking the limit as $D \to 4$. This allows the GB term to contribute nontrivially to gravitational dynamics, leading to what is known as 4D Einstein-Gauss-Bonnet (EGB) gravity. This regularization process was initially considered by Tomozawa \cite{Tomozawa:2011gp}, who included finite one-loop quantum corrections to Einstein gravity. Wormhole solutions were also investigated in 4D EGB gravity. For instance, Ref.\cite{Jusufi:2020yus} discussed wormhole solutions in 4D EGB gravity for isotropic and anisotropic matter sources. In Ref.\cite{Kokubu:2020lxs,Zhang:2020kxz}, the possibility for eternal existence of thin-shell wormholes has been investigated. Moreover, the stability of thin-shell wormholes \cite{Godani:2022jwz} and Yukawa-Casimir wormholes \cite{Shweta:2023hvm} has been investigated. Moreover, wormholes within galactic dark matter halos has been recently explored in Ref.\cite{Hassan:2024xyx}.

Among these fascinating structures, charged traversable wormholes stand out due to their additional complexity and potential implications. A charged wormhole, influenced by electromagnetic fields, must satisfy the Einstein-Maxwell equations, which integrate the principles of general relativity with electromagnetism. Therefore, the work of Ref.\cite{Jusufi:2020yus} can be extended to include also the electromagnetic field as an additional source. Charged wormholes appear in many theories of gravity, often providing intriguing solutions that combine both electromagnetic and gravitational fields to explore the stability, structure, and physical properties of these exotic spacetime configurations. Regarding charged solutions, the authors of Ref.\cite{Mehdizadeh:2018smu} study wormhole solutions in Einstein-Cartan theory (ECT) in the presence of electric charge; while, the exact solutions of wormhole with extra fields such as scalar field and electric charge have been addressed in Ref.\cite{Kim:2001ri}.

This work is organized as follows: We give a short review on charged EGB gravity \& wormhole solutions in Sec.\ref{sec2}. Subsequently, the embedding diagrams of the wormholes are illustrated in Sec.\ref{sec3}. In Sec.\ref{energy}, we study the energy conditions of the charge wormhole solutions. We also use the stability analysis to examine whether the wormhole solutions are in equilibrium in Sec.\ref{sec4}. Our conclusions are drawn in the last section.

\section{A short review on charged EGB gravity \& Wormhole solutions}\label{sec2}
Recently, 4D Einstein–Gauss–Bonnet (4DEGB) gravity has attracted considerable attention. Initially, we introduce Lovelock's theorem as a foundational concept, focusing particularly on Gauss–Bonnet terms within the gravity action, which are extensively reviewed in Ref.\cite{Fernandes:2022zrq}. However, our discussion closely follows the approach outlined in Ref.\cite{Glavan:2019inb}. Starting with the standard Einstein-Gauss-Bonnet action without a cosmological constant, we incorporate the Maxwell field, modifying the Gauss-Bonnet coupling constant $\alpha\rightarrow \alpha/(D-4)$, as described in Ref.\cite{Fernandes:2020rpa}.

\begin{equation}\label{action}
	S=\int d^{D}x\sqrt{-g}\left[ \frac{1}{16 \pi}R +\frac{\alpha}{D-4} \mathcal{L}_{\text{GB}} -
\frac{1}{4}F_{\mu\nu}F^{\mu\nu}\right]
+\mathcal{S}_{\text{matter}},
\end{equation}
where $g$ denotes the determinant of the metric $g_{\mu\nu}$ and $\alpha$ is the Gauss-Bonnet coupling coefficient with
dimension $[\rm length]^2$ and we assume that it is positive here, $F_{\mu\nu}= \partial_{\mu}A_{\nu} - \partial_{\nu}A_{\mu}$ is the usual Maxwell tensor. The term $\mathcal{L}_{\text{GB}}$ is the Lagrangian defined by
\begin{equation}
	\mathcal{L}_{\text{GB}}=R^{\mu\nu\rho\sigma} R_{\mu\nu\rho\sigma}- 4 R^{\mu\nu}R_{\mu\nu}+ R^2\label{GB}.
\end{equation}
In Eq.(\ref{action}), $S_{\rm matter}$ represents the matter fields within the theory. By varying this action with respect to the metric $g_{\mu \nu}$ and the field strength tensor $F_{\mu\nu}$, the following field equations can be derived \cite{EslamPanah:2020hoj}:
\begin{eqnarray}\label{GBeq}
	G_{\mu\nu}+\frac{\alpha}{D-4} H_{\mu\nu} &=& 8 \pi T_{\mu\nu},\\
 \nabla_{\mu}F_{\mu\nu}&=&0\,.
\end{eqnarray}
where $G_{\mu\nu}$ is the usual Einstein tensor, $ T_{\mu\nu}=T^{\rm m}_{\mu\nu}+T^{\rm em}_{\mu\nu}$, $ T^{\rm m}_{\mu\nu}= -\frac{2}{\sqrt{-g}}\frac{\delta\left(\sqrt{-g}\mathcal{S}_m\right)}{\delta g^{\mu\nu}}$ is the energy momentum tensor of matter, while $H_{\mu\nu}$ and $T^{\rm em}_{\mu\nu}$ read
\begin{eqnarray}
	H_{\mu\nu}&=&2\Bigr( R R_{\mu\nu}-2R_{\mu\sigma} {R}{^\sigma}_{\nu} -2 R_{\mu\sigma\nu\rho}{R}^{\sigma\rho} - R_{\mu\sigma\rho\delta}{R}^{\sigma\rho\delta}{_\nu}\Bigl)-\frac{1}{2}g_{\mu\nu}\mathcal{L}_{\text{GB}},\label{FieldEq}\\
    T^{\rm em}_{\mu\nu}&=&\frac{1}{4\pi}\Big(F^{\lambda}_{\mu}F_{\nu\lambda}-\frac{1}{4}g_{\mu\nu}F^{\alpha\beta}F_{\alpha\beta}\Big)\,.
\end{eqnarray}
Here, $R$ denotes the Ricci scalar, $R_{\mu\nu}$ the Ricci tensor, $H_{\mu\nu}$ is the Lanczos tensor, and $R_{\mu\sigma\nu\rho}$ represents the Riemann tensor. It is straightforward to demonstrate that the electromagnetic stress–energy tensor, $T^{\rm em}_{\mu\nu}$, is traceless. To construct a wormhole within the framework of charged 4D Einstein-Gauss-Bonnet (EGB) gravity, we consider the general static, spherically symmetric $D$-dimensional metric discussed in \cite{Mehdizadeh:2015jra,Zubair:2023wjh}, given by
\begin{equation}
ds^2=-e^{2\Phi(r)}dt^2+\frac{dr^2}{1-\frac{b(r)}{r}}+r^2d\Omega_{D-2}^2.\label{metric}
\end{equation} 
where
\begin{equation}
d\Omega^2_{D-2} = d\theta^2_1 + \sum^{D-2}_{i=2}\prod^{i-1}_{j=1}
\sin^{2}\theta_j\;d\theta^2_i \;,\notag
\end{equation}
where the function $\Phi(r)$ represents the redshift function of an infalling body, which must be finite everywhere to prevent the presence of an event horizon. Conversely, $b(r)$ denotes the spatial shape function of the wormhole geometry, determining its shape in the embedding diagram \cite{Mehdizadeh:2015jra}. It is crucial that $b(r)$ satisfies the boundary condition $b(r=r_0) = r_0$ at the throat $r_0$, where $r_0 \leq r \leq \infty$. Moreover, to ensure the traversability of the wormhole, the function $b(r)$ must adhere to the \textit{flaring-out} condition, which is derived from the embedding calculation and is expressed as follows:
\begin{equation}\label{a12}
\frac{b(r)-rb^{\prime}(r)}{b^2(r)}>0.
\end{equation}
We consider an anisotropic fluid for the matter source defined by the stress energy tensor
\begin{eqnarray}
T^\nu_i &=& (\rho+p_t)u^\nu u_i + p_t g^\nu_i + (p_r-p_t)\chi_i \chi^\nu, \label{eq8}
\end{eqnarray}
where, $u_\nu$ represents the 4-velocity and $\chi_\nu$ denotes the unit spacelike vector in the radial direction, where $\rho(r)$ stands for the energy density, and $p_r(r)$ and $p_t(r)$ denote the radial and transverse pressures, respectively. The field equations in Einstein-Gauss-Bonnet (EGB) gravity can be derived using the metric given in Eq. (\ref{metric}), as detailed in \cite{Zubair:2023wjh}. The corresponding electromagnetic potential is expressed as:
\begin{eqnarray}
A^\mu =-\frac{q}{r}dt\delta^{\mu}_{t}\,. \label{aa}
\end{eqnarray}
The only non-vanishing components of the Faraday tensor take the form:
\begin{eqnarray}
F_{01}=-F_{10}=-\frac{q}{r^{2}}, \label{f01}
\end{eqnarray}
where $q$ denotes an integration constant associated with the electric charge. By employing the metric given in Eq. (\ref{metric}) along with the stress tensor from Eq. (\ref{eq8}), in the limit as $D \to 4$, the components of the field equations (\ref{GBeq}) can be expressed as:
\begin{eqnarray}
&& 8\pi \Big(\rho(r)+\frac{q^{2}}{8\pi r^{4}}\Big)= \frac{b(r)}{r^3}\Big(1-\frac{\alpha  b(r)}{r^3}\Big)+\frac{1}{r^2}\left(\frac{2 \alpha  b(r)}{r^3}+1\right) \left(b'(r)-\frac{b(r)}{r}\right)\,,\label{DRE1}\\
&& 8\pi \Big(p_r(r)-\frac{q^{2}}{8\pi r^{4}}\Big)= -\frac{b(r) \left(1-\frac{\alpha  b(r)}{r^3}\right)}{r^3}+\frac{2}{r}\left(1-\frac{b(r)}{r}\right) \left(\frac{2 \alpha  b(r)}{r^3}+1\right)\Phi'(r) \label{DRE2}\,,
\end{eqnarray}
and
\begin{eqnarray}\label{DRE3}
8\pi \Big(p_t(r)+\frac{q^{2}}{8\pi r^{4}}\Big) &=& -\frac{\alpha  b(r)^2}{r^6}+\frac{2 \alpha }{r^4}\left(1-\frac{b(r)}{r}\right) \left(b(r)-r b'(r)\right) \Phi '(r)\nonumber\\&+&\left(1-\frac{b(r)}{r}\right) \left(\frac{2 \alpha  b(r)}{r^3}+1\right) \left(\frac{\left(b(r)-r b'(r)\right) \Phi '(r)}{2 r (r-b(r))}+\Phi ''(r)-\Phi '(r)^2\right)\nonumber\\&+&\left(1-\frac{b(r)}{r}\right) \left(1-\frac{2 \alpha  b(r)}{r^3}\right) \left(\frac{b(r)-r b'(r)}{2 r^2 (r-b(r))}-\frac{\Phi '(r)}{r}\right)\,, \label{DRE3}
\end{eqnarray}
where the prime denotes a derivative with respect to the
radial coordinate $r$. In this context, we have five unknown functions of $r$, i.e., $\rho(r)$, $p_r(r)$, $p_t(r)$, $b(r)$ and $\Phi(r)$. In the following, we examine wormhole solutions in 4D Einstein-Gauss-Bonnet
Gravity.

\subsection{Isotropic solution}
To simplify the problem, we consider a constant redshift function, $\Phi(r) = \text{const.}$, implying a wormhole solution with zero tidal force. This choice simplifies calculations and yields intriguing exact wormhole solutions. Recent studies \cite{Fernandes:2020nbq,Hennigar:2020lsl} have demonstrated that by taking the trace of the field equations (3), one obtains the following simple form:
\begin{equation}\label{n12}
    R+\frac{\alpha}{2}\mathcal{L}_{\text{GB}}=-8\pi T,
\end{equation}
where the trace $T=T^{\mu}_{ \nu}$. Using the metric form (\ref{metric}) for isotropic fluid matter we can use the relation $
    p_r(r)=p_t(r) =\omega \rho(r)
$  obtain the following condition 
\begin{equation}
    \frac{2 b'(r)}{r^2}+8\pi \left[-\rho(r)+3 \omega \rho(r) \right]=0.\label{trace}
\end{equation}
Utilizing Eq.(\ref{trace}) and  (\ref{DRE1}), we obtain
a differential equation for the shape function as
\begin{eqnarray}
    \frac{r^4 (3 \omega +1) b'(r)}{3 \omega -1}+\alpha  b(r) \left(2 r b'(r)-3 b(r)\right)-q^2 r^2=0.
\end{eqnarray}
Solving the above equation for $b(r)$
which has the following form
\begin{equation}\label{brs}
    b(r)=-\frac{r^3 (3 \omega +1)}{2 \alpha  (3 \omega -1)}\Bigg(1\pm \sqrt{1+\frac{4\alpha{\cal A}}{r^3 r_{0}}}\Bigg) ,
\end{equation}
where we have defined a new parameter ${\cal A}$ as
\begin{equation}
    {\cal A}=\Big(\frac{q^2 (3 \omega -1)^2 (r-r_{0})}{r(3 \omega +1)^2}+ \frac{(3 \omega -1) \left(-\alpha +3 \omega  \left(\alpha +r_{0}^2\right)+r_{0}^2\right)}{(3 \omega +1)^2}\Big)/,,
\end{equation}
where we have applied the condition $b(r=r_0)=r_0$, noting that when $q=0$, our results reduce to those presented in Ref.\cite{Jusufi:2020yus}. It's important to emphasize that the solution given by Eq.(\ref{brs}) is valid only for $\omega \neq \pm 1/3$. The $\pm$ sign in Eq.(\ref{brs}) denotes two distinct branches of solutions. In studies such as \cite{Boulware:1985wk,Wheeler:1985nh}, it has been demonstrated that Einstein-Gauss-Bonnet (EGB) black holes with the $+$ branch are unstable, exhibiting a ghost graviton degree of freedom, whereas those with the $-$ branch are stable and free of ghosts. In our scenario, in the limit $\alpha \to 0$, the positive branch leads to
\begin{equation}
    \frac{b(r)}{r}=-\frac{r^2(3 \omega+1)}{(3\omega-1)\alpha}-\frac{r_0}{r}-\frac{q^2 (3 \omega -1) (r-r_{0})}{r^2 r_{0} (3 \omega +1)} \ldots
\end{equation}
which represents a wormhole solution within de-Sitter/anti-de Sitter spacetimes, depending on the sign of $\alpha$. Conversely, in the limit as $\alpha \to 0$, the negative branch transitions to
\begin{equation}\label{viab}
    \frac{b(r)}{r}=\frac{r_0}{r}+ \frac{q^2 (3 \omega -1) (r-r_{0})}{r^2 r_{0} (3 \omega +1)}+\ldots,
\end{equation}
 and the standard Morris-Thorne wormhole is obtained when the charge vanishes. Henceforth we restrict to the negative branch in which case the 4D EGB wormhole metric reads 
\begin{equation}\label{metricw}
    ds^2=-dt^2+\frac{dr^2}{1+\frac{r^2(3\omega+1)}{2\alpha (3 \omega-1)}\left( 1 \pm \sqrt{1+\frac{4 \alpha \mathcal{A} }{ r^3 r_0 }  }    \right) }  +r^2 d\Omega^2_2,
\end{equation}
where the unit sphere line element is given by $d\Omega_2^2 = d\theta^2 + \sin^2\theta , d\phi^2$. It's important to note that the above analysis holds true for the interval $\omega \geq -1/3$ and $\omega \neq 1/3$. For values $\omega < -1/3$, the sign of the solutions reverses.
\begin{figure}[t]
    \centering
    \includegraphics[width = 8 cm]{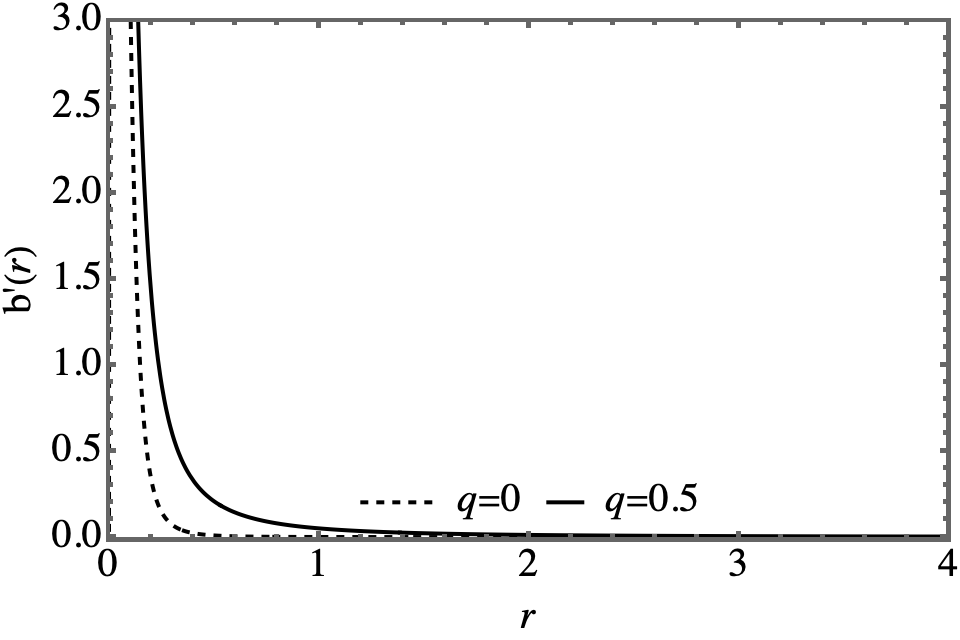}
    \includegraphics[width = 8.3 cm]{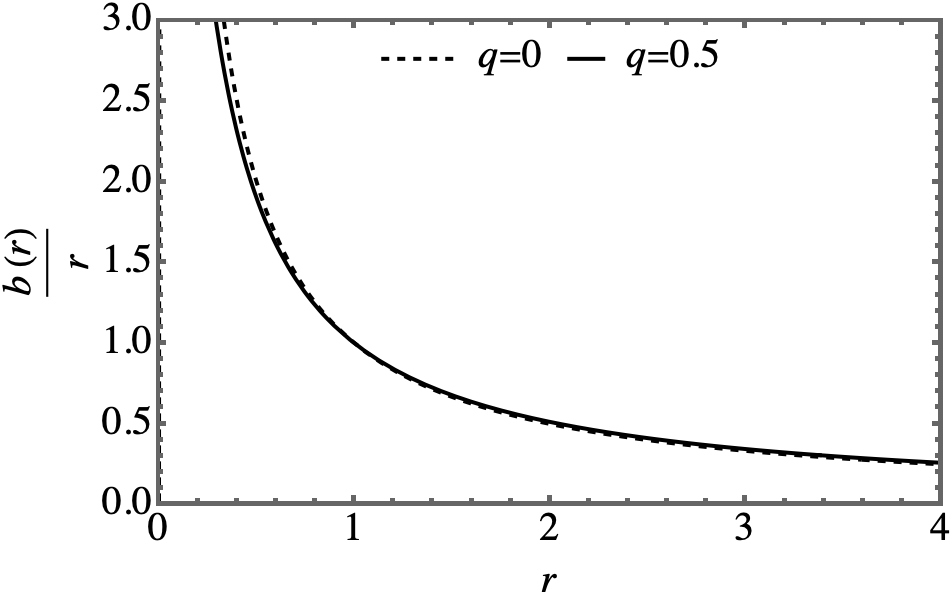}
    \caption{Figures show the behaviors of $b(r)$ and $b'(r)$ as a function of $r$ for our exact isotropic wormhole. We have used the constants $\alpha = 0.001$ and $r_0 = 1$ along with $\omega = 0.5$.}
    \label{biso}
\end{figure}
To obtain a TW we need to compute the redshift function. We have to impose the EoS $p_{r}(r) = \omega \rho(r)$. From Eq.(\ref{DRE1}) \& (\ref{DRE2}), one finds
\beq
\Phi'(r) = \frac{r^4 \omega  b'(r)+2 \alpha  r \omega  b(r) b'(r)+r^3 b(r)-3 \alpha  \omega  b(r)^2-\alpha  b(r)^2-q^2 r^2}{2 r (r-b(r)) \left(2 \alpha  b(r)+r^3\right)}\,.\label{Phiiso}
\eeq
Substituting $b(r)$ from Eq.(\ref{viab}), we come up with
\beq\label{Phidvi}
\Phi'(r) &=&\frac{3 q^2 r \omega -8 q^2 r_0 \omega +q^2 (-r)+3 r r_0^2 \omega +r r_0^2}{2 r \left(r-r_0\right) \left(-3 q^2 \omega +q^2+3 r r_0 \omega +r r_0\right)}\nonumber\\&&-\frac{\alpha}{2 \left(r^5 \left(r-r_0\right) r_0 (3 \omega +1) \left(-3 q^2 \omega +q^2+3 r r_0 \omega +r r_0\right)\right)}\times\nonumber\\&&\times\Big(\big(3 q^2 r \omega -3 q^2 r_0 \omega +q^2 (-r)+q^2 r_0+3 r r_0^2 \omega +r r_0^2\big) \nonumber\\&&\quad\big(9 q^2 r \omega ^2-15 q^2 r_0 \omega ^2+6 q^2 r \omega -14 q^2 r_0 \omega -3 q^2 r+q^2 r_0+9 r r_0^2 \omega ^2+12 r r_0^2 \omega +3 r r_0^2\big)\Big)\nonumber\\&&+O\left(\alpha ^2\right)\,.
\eeq
Let us consider solely the first term, we have
\beq\label{Phidvi}
\Phi'(r) \simeq\frac{3 q^2 r \omega -8 q^2 r_0 \omega +q^2 (-r)+3 r r_0^2 \omega +r r_0^2}{2 r \left(r-r_0\right) \left(-3 q^2 \omega +q^2+3 r r_0 \omega +r r_0\right)}\,.
\eeq
We choose $\omega$ such that $\Phi'(r)=0$ to avoid the appearance of a horizon. This can be seen if
\beq\label{omq}
\omega \simeq \frac{r \left(q^2-r_0^2\right)}{3 q^2 r-8 q^2 r_0+3 r r_0^2}\,.
\eeq
We next assume the small charge approximation and expand (\ref{omq}) to obtain
\beq\label{omq1}
\omega \simeq -\frac{1}{3}+\frac{2 q^2 (3 r-4 r_{0})}{9 r r_{0}^2}+{\cal O}\left(q^3\right)\,.
\eeq
Interestingly, it is worth mentioning here that due to the present of a charge contribution, $\omega$ naturally deviates from $-\tfrac{1}{3}$.

\subsection{Anisotropic fluid scenario}
We shall argue that there are two ways to find an anisotropic solution. Taking into account Eqs. (\ref{DRE2}-\ref{DRE3}) along with an interesting EoS \cite{Moraes:2017dbs},
\begin{equation}\label{n23}
p_t(r)=\omega_t p_r(r),
\end{equation}
leads to the following expression
\begin{eqnarray}\notag
r \left(r^3-2 \alpha  b(r)\right) b'(r)+b(r) \left(2 \alpha  (\omega_{t} +2) b(r)-r^3 (2 \omega_{t} +1)\right)+2 q^2 r^2 (\omega_{t} +1)=0,
\end{eqnarray}
where the EoS parameter $\omega_t$ and the redshift function $\Phi(r)$ are constants. Solving the last differential equation we get the following shape function 
\begin{equation}\label{n25}
    b(r)=\frac{r^3}{2 \alpha}\left[1\pm\sqrt{\frac{r^4-4 \alpha  q^2}{r^4}+4 \alpha ^2 c_1 r^{2 \omega_{t} -2}}    \right],
\end{equation}
where $c_1$ is a constant of integration, which can be fixed using the condition $b(r=r_0)=r_0$, Then the shape function modifies to
\begin{equation}\label{n26}
    b(r)=\frac{r^3}{2 \alpha} \left[1 \pm \sqrt{1-\frac{4 \alpha  q^2}{r^4}+\frac{4 \alpha  r^{2 \left(\omega _t-1\right)}}{r_{0}^{2 (\omega_{t} +1)}}\left(\alpha +q^2-r_{0}^2\right)}  \right].
\end{equation}

Similarly for the $+$ve branch sign in the limit $\alpha \to 0$, we obtain
\begin{equation}
    \frac{b(r)}{r}=\frac{r^2}{\alpha }+\frac{r^{2\omega_t}}{r_{0}^{2(\omega_{t} +1)}}\left(q^2-r_{0}^2\right)-\frac{q^2}{r^2}+...
\end{equation}
which is a wormhole solution in a de-Sitter/ anti-de Sitter spacetimes depending on the sign of $\alpha$. On the other hand, in the limit $\alpha \to 0$, the  $-$ve  goes over
\begin{equation}
    \frac{b(r)}{r}=-\frac{r^{2\omega_t}}{r_{0}^{2(\omega_{t} +1)}}\left(q^2-r_{0}^2\right)+\frac{q^2}{r^2}+ \ldots,
\end{equation}
 which reduces to the standard Morris-Thorne wormhole. Henceforth, we restrict to the $-$ve branch for anisotropic fluid in the   4D EGB  takes the form
\begin{equation}\label{eq39}
    ds^2=-dt^2+\frac{dr^2}{1-\frac{r^2}{2 \alpha} \left[1 - \sqrt{1-\frac{4 \alpha  q^2}{r^4}+\frac{4 \alpha  r^{2 \left(\omega_t-1\right)}}{r_{0}^{2 (\omega_{t} +1)}}\left(\alpha +q^2-r_{0}^2\right)}\right]}+r^2d\Omega_2^2.
\end{equation}
\begin{figure}[t]
    \centering
    \includegraphics[width = 8.1 cm]{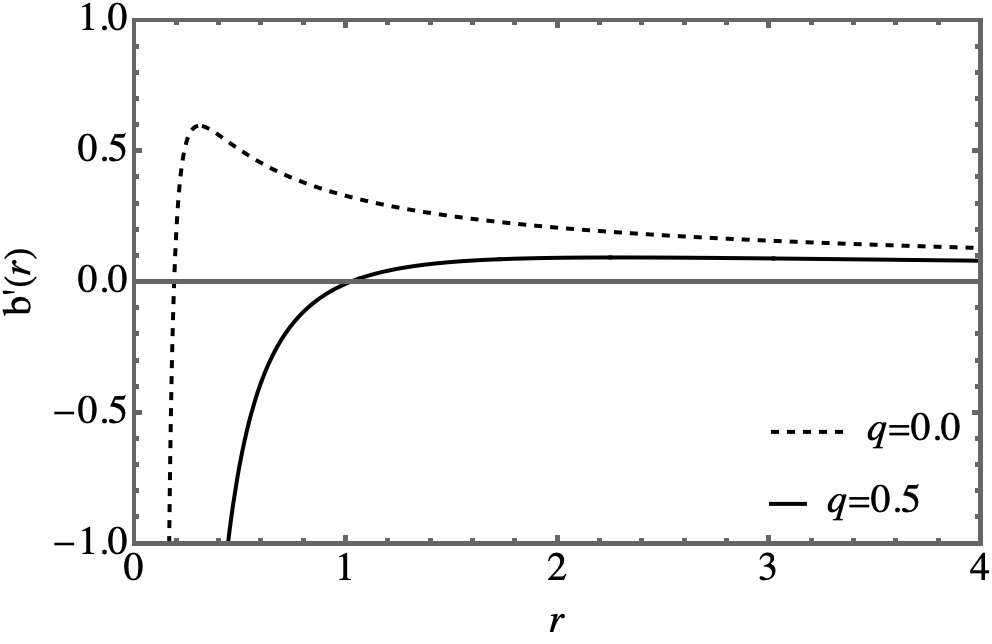}
    \includegraphics[width = 8.1 cm]{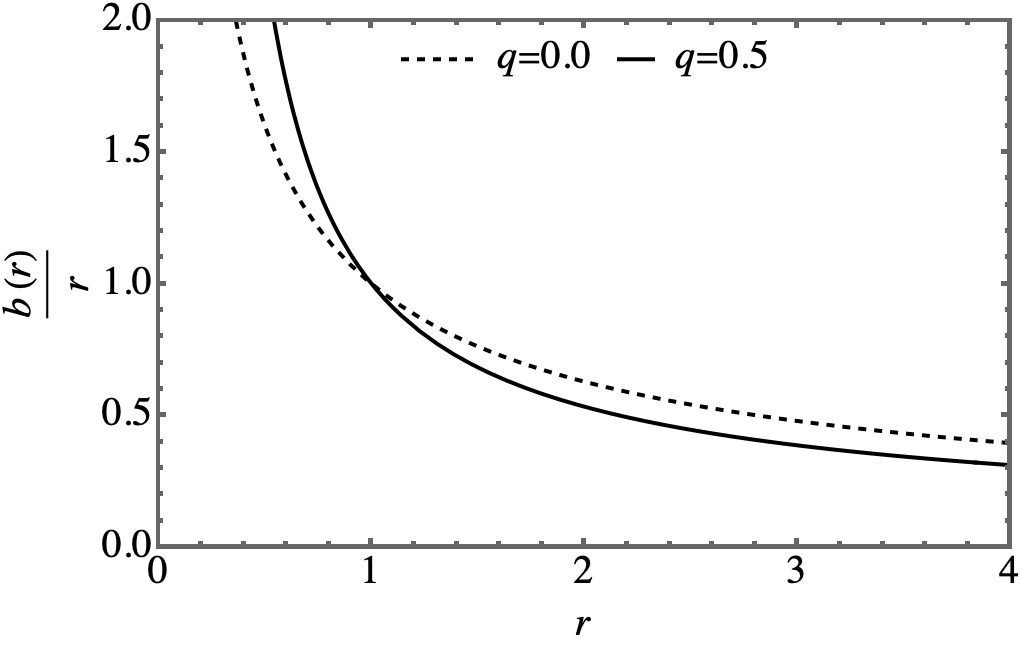}
    \caption{Figures show the behaviors of $b(r)$ and $b'(r)$ as a function of $r$ for an isotropic wormhole. We have used the constants $\alpha = 0.001$ and $r_0 = 1$ along with $\omega_t = -1/3$.}
    \label{biso}
\end{figure}

\subsection{Model with a specific energy density}
To determine the shape function, one can consider a specific energy density profile. Here, we will adopt the following energy density \cite{Lobo:2012qq}:
\begin{equation}
    \rho(r)=\rho_{0}\left(\frac{r_{0}}{r}\right)^{\beta },
\end{equation}
where $\beta$ and $\rho_0$ are constants. Solving the Eq. (\ref{DRE1}), we obtain the following shape function
\begin{equation}\label{eq24}
    b(r)=-\frac{r^3}{\alpha}\left(1 \pm \sqrt{1+4 \alpha  \left(-\frac{8 \pi \rho_{0} \left(\frac{r_{0}}{r}\right)^{\beta }}{\beta -3}+\frac{-q^2+\alpha  c_1 r}{r^4}\right)}   \right),
\end{equation}
in which $c_{1}$ is a constant of integration. Using the condition $b(r=r_0)=r_0$, we obtain
\begin{equation}
    c_{1}=\frac{\alpha +q^2+r_{0}^2}{\alpha r_{0}}+\frac{8 \pi  \rho_{0} r_{0}^4}{\alpha  (\beta -3) r_{0}}.
\end{equation}
Substituting the above expression into the Eq. (\ref{eq24}), one obtains the two different branches of the solution, which are
\begin{equation}\label{Anla0}
     b(r)=-\frac{r^3}{\alpha}\left(1 \pm \sqrt{1-\frac{4 \alpha  q^2}{r^4}+\frac{4 \alpha}{r^3 r_{0}}\left(\alpha +q^2+\frac{8 \pi  \rho_{0} r_{0}^4}{\beta -3}+ r_{0}^2\right)-\frac{32 \pi  \alpha \rho_{0}}{\beta -3}\left(\frac{r_{0}}{r}\right)^{\beta }}   \right).
\end{equation}
The branch with the negative sign is stable and we select it as the physical solution, under the condition $\beta \geq 4$. However, when $\beta = 3$, there appears to be a singularity. It's important to note that there is a third case where the sign of the solution flips, specifically when $\beta = 1$ or $2$. The negative branch solutions for $\beta \geq 4$ are asymptotically flat; for instance, choosing $\beta = 4$ illustrates this behavior in the limit $\alpha \to 0$, where the negative branch gives
\begin{equation}
    \frac{b(r)}{r}=-\frac{q^2}{r^2}\Big(1-\frac{r}{r_{0}}\Big)-\frac{8 \pi \rho_{0} r^4 \left(\frac{r_{0}}{r}\right)^{\beta }+r r_{0} \left(\beta +8 \pi \rho_{0} r_{0}^2-3\right)}{(\beta -3) r^2}+ \ldots
\end{equation}
 On the other hand, in the limit $\alpha \to 0$, the  positive branch goes over
\begin{equation}
    \frac{b(r)}{r}=-\frac{r^2}{\alpha}+\frac{q^2}{r^2}\Big(1-\frac{r}{r_{0}}\Big)+\frac{8 \pi \rho_{0} r^4 \left(\frac{r_{0}}{r}\right)^{\beta }-r r_{0} \left(\beta +8 \pi \rho_{0} r_{0}^2-3\right)}{(\beta -3) r^2}+ \ldots,
\end{equation}
corresponds to the wormhole solution in de-Sitter/anti-de Sitter spacetimes. Finally, it can be verified that for $\beta=1,2$, our solutions (\ref{Anla0}) are not asymptotically flat. Using the shape function (\ref{Anla0}), the wormhole metric is defined by the expression:
\begin{equation}
    ds^2=-dt^2+\frac{dr^2}{1+\frac{r^2}{\alpha}\left(1- \sqrt{1+\zeta}  \right)}+r^2d\Omega^2_2,
\end{equation}
where
\begin{equation}
    \zeta=-\frac{4 \alpha  q^2}{r^4}+\frac{4 \alpha}{r^3 r_{0}}\left(\alpha +q^2+\frac{8 \pi  \rho_{0} r_{0}^4}{\beta -3}+ r_{0}^2\right)-\frac{32 \pi  \alpha \rho_{0}}{\beta -3}\left(\frac{r_{0}}{r}\right)^{\beta }, \notag
\end{equation}
under the condition $\beta \geq 4$. However, as previously noted, caution is required since for $\beta=1,2$, the sign of the solution flips. As a special case, one can consider the limit $\beta \to 0$, where the energy density becomes constant, i.e., $\rho=\rho_0$.  
\begin{figure}[t]
    \centering
    \includegraphics[width = 8.1 cm]{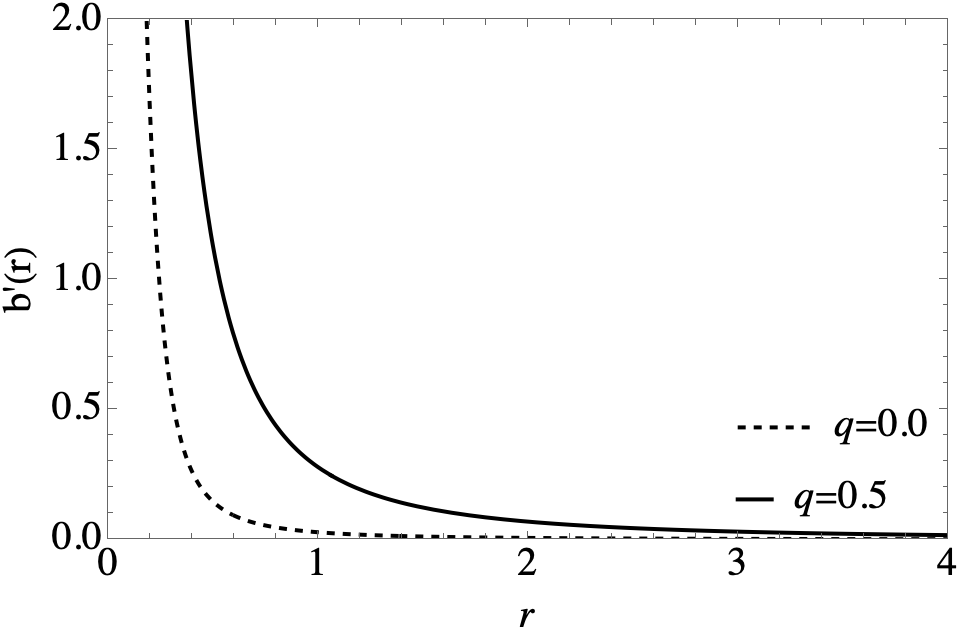}
    \includegraphics[width = 8.1 cm]{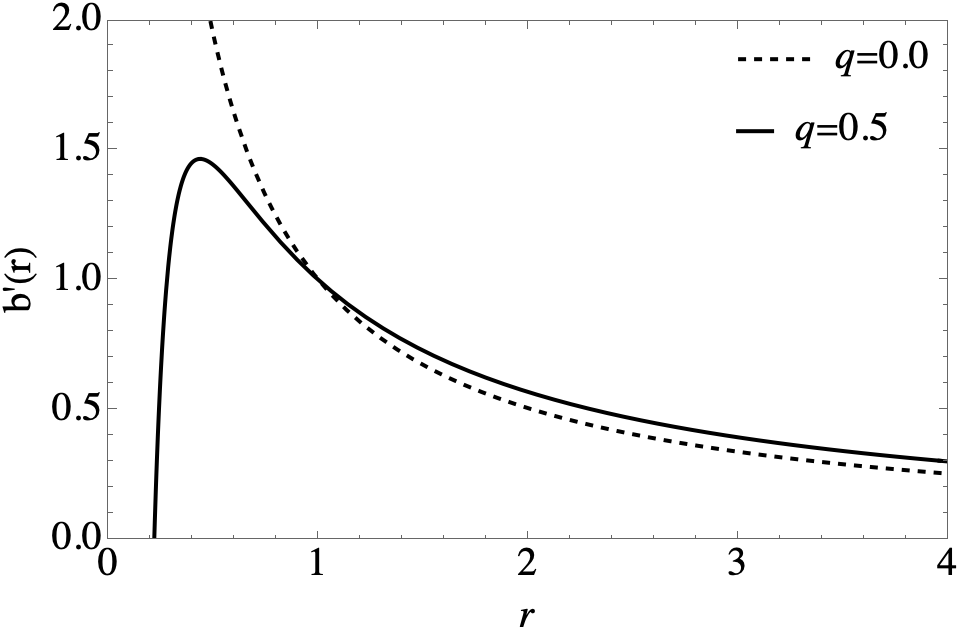}
    \caption{Figures show the behaviors of $b(r)$ and $b'(r)$ as a function of $r$ for a model with a specific energy density. We have used the constants $\alpha = 0.001,\, \beta = 4,\, \rho_{0}= 0.01,\, q= 0.5,\, r_{0}= 1$.}
    \label{biso}
\end{figure}

\section{Embedding diagram}
\label{sec3}
\begin{figure}[!h]	
\begin{center}		
\includegraphics[width=9.1cm]{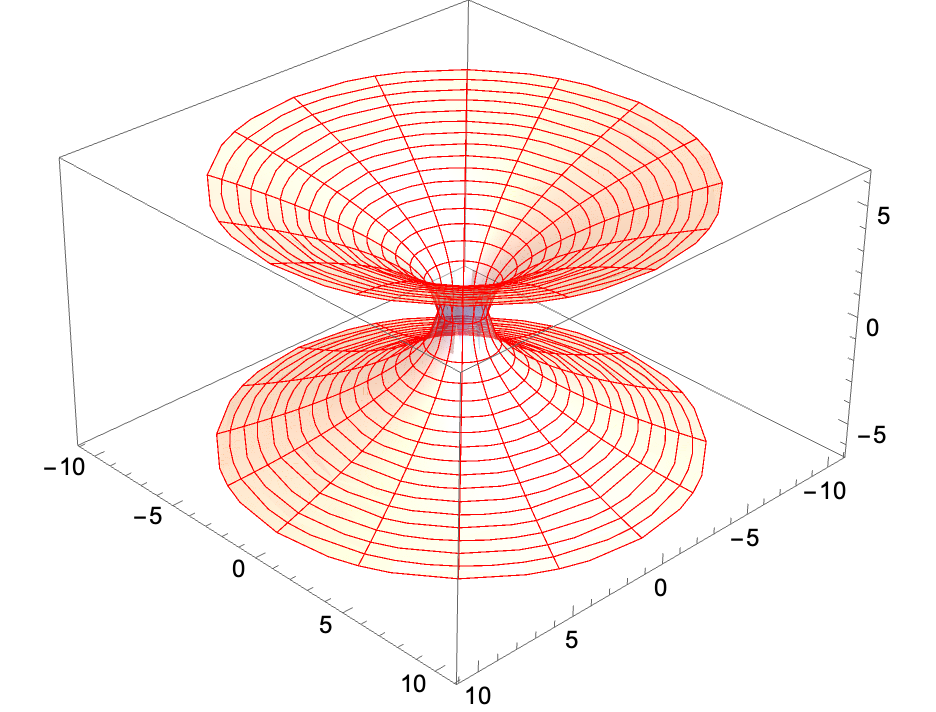}
\includegraphics[width=7.1cm]{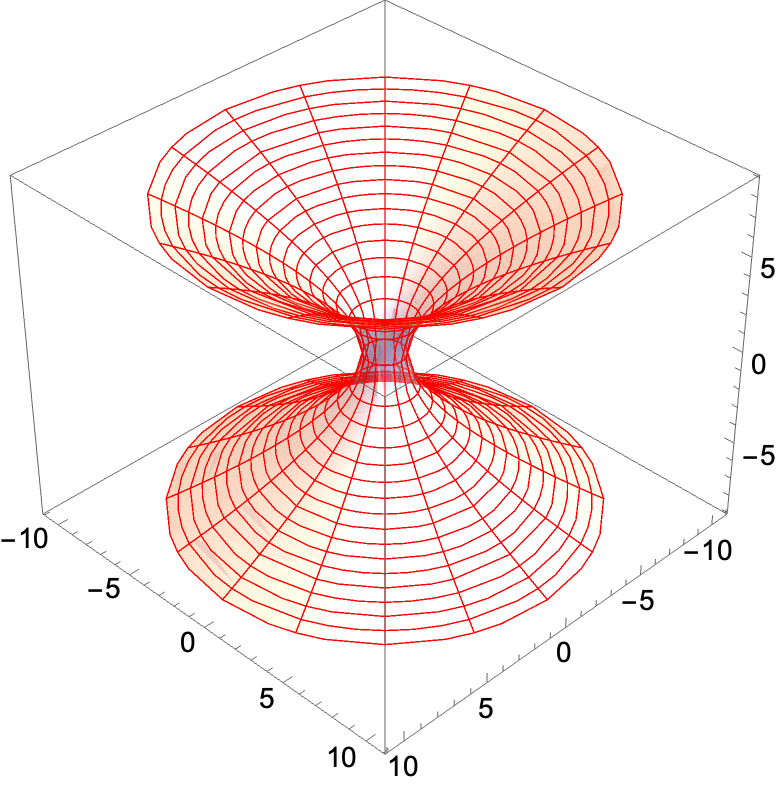}
\caption{The embedding diagram of wormholes geometry along the equatorial plane ($t = {\rm const.}, \theta =\pi/2$). The specific case of a constant redshift $\Phi'(r)=0$ with $\alpha = 0.001, q=0.3$ and $r_{0}=1$, we have drawn embedding diagram for isotropic case and an anisotropic wormhole illustrated on the left and right panel, respectively. We have considered the numerical values of $\omega=0.5$ and $\omega_{t} = -1/3$ and $\beta=4$.}
\label{gmod3}
\end{center}
\end{figure}

In this subsection, we examine the embedding diagrams to depict the charged wormhole solutions in 4D Einstein-Gauss-Bonnet Gravity, focusing on an equatorial slice $\theta=\pi/2$ at a fixed moment in time $t={\rm constant}$. To achieve this, we consider the metric expressed as
\begin{eqnarray}
ds^{2}=\frac{dr^{2}}{1-\frac{b(r)}{r}}+r^{2}d\phi^{2}\,.
\end{eqnarray}
We embed the metric (\ref{metric}) into three-dimensional Euclidean space to visualize this slice. The spacetime can then be described using cylindrical coordinates as
\begin{eqnarray}
ds^{2}=dz^{2}+dr^{2}+r^{2}d\phi^{2}\,.
\end{eqnarray}
From the above equations, we can write
\begin{eqnarray}
\frac{dz}{dr}=\pm \sqrt{\frac{r}{r-b(r)}-1}\,,
\end{eqnarray}
where $b(r)$ is given by Eq.(\ref{brs}), Eq.(\ref{n26}) and Eq.(\ref{Anla}). However, in this section, we only focus on exploring the geometrical properties of these matrices Eq.(\ref{brs}) and Eq.(\ref{n26}) via the embedding diagram. Invoking numerical techniques allows us to illustrate the wormhole shape given in Fig.\ref{gmod3}.

\section{Energy conditions}
\label{energy}
In this section, we provide a detailed description of the energy conditions, which are sets of inequalities based on the energy-momentum tensor. Specifically, we begin by finding wormhole solutions for the weak energy condition (WEC), defined as $T_{\mu\nu}U^{\mu}U^{\nu}$, where $U^{\mu}$ is a timelike vector. For the given diagonal energy-momentum tensor, the WEC implies:
\beq
 \rho(r) \geq 0\quad{\rm and}\quad \rho(r)+P_{i}(r)\geq 0\,.
 \eeq
 Next, the null energy condition (NEC) is defined by $T_{\mu\nu}k^{\mu}k^{\nu}$, where $k^{\mu}$ is a null vector. For the given diagonal energy-momentum tensor, the NEC implies:
\beq
 \rho(r) \geq 0\quad{\rm and}\quad \rho(r)+P_{i}(r)\geq 0\,.
 \eeq
 The strong energy condition (SEC) asserts that $T_{\mu\nu}U^{\mu}U^{\nu}\geq \tfrac{1}{2}Tg_{\mu\nu}U^{\mu}U^{\nu}$ for any timelike vector $U^{\mu}$, ensuring that gravity is attractive. The SEC implies:
\beq
 \rho(r) +\sum_{i} P_{i}(r)\geq 0\quad{\rm and}\quad \rho(r)+P_{i}(r)\geq 0\,.
 \eeq
 Note that while the WEC or SEC imply the NEC, any violation of the NEC also indicates a violation of the SEC and WEC. In the following, we present the energy conditions for all specific cases.
 
\subsection{Isotropic solution}
We begin with the isotropic solution. Using the key Eqs.(\ref{DRE1}-\ref{DRE3}), we can compute $\rho$ and $p_{r}$ to obtain
\beq
 \rho &=&B^{-1}\Big(3 (A-1) r_0 r^4 (3 \omega +1)^2-6 \alpha  r (3 \omega -1) \big(-\alpha +3 \omega  \big(\alpha +q^2+r_0^2\big)\nonumber\\&&\quad\quad\quad\quad\quad\quad\quad\quad\quad -q^2+r_0^2\big)+4 \alpha  q^2 r_0 (1-3 \omega )^2\Big)\,,\\
 p_{r}&=&B^{-1}\Big(3 (A-1) r_0 r^4 (3 \omega +1)^2-6 \alpha  r (3 \omega -1) \big(-\alpha +3 \omega  \big(\alpha +q^2+r_0^2\big)\nonumber\\&&\quad\quad\quad\quad\quad\quad\quad\quad\quad -q^2+r_0^2\big)+4 \alpha  q^2 r_0 (1-3 \omega )^2\Big)\,,
 \eeq
where
\beq
A&\equiv& \sqrt{1+\frac{4 \alpha  (3 \omega -1) \left(q^2 \left(r-r_0\right) (3 \omega -1)+r \left(-\alpha +3 \omega  \left(\alpha +r_0^2\right)+r_0^2\right)\right)}{r^4 r_0 (3 \omega +1)^2}},\\B&\equiv&8 \pi  \alpha  A r^4 r_0 (1-3 \omega )^2 (3 \omega +1)\,.
 \eeq
In particular, it is evident from Fig.(\ref{energyiso}) that for $\alpha>0$, the NEC, and consequently the WEC, are violated at the throat due to the flaring-out condition.
 
\begin{figure}[t]
    \centering
    \includegraphics[width = 8 cm]{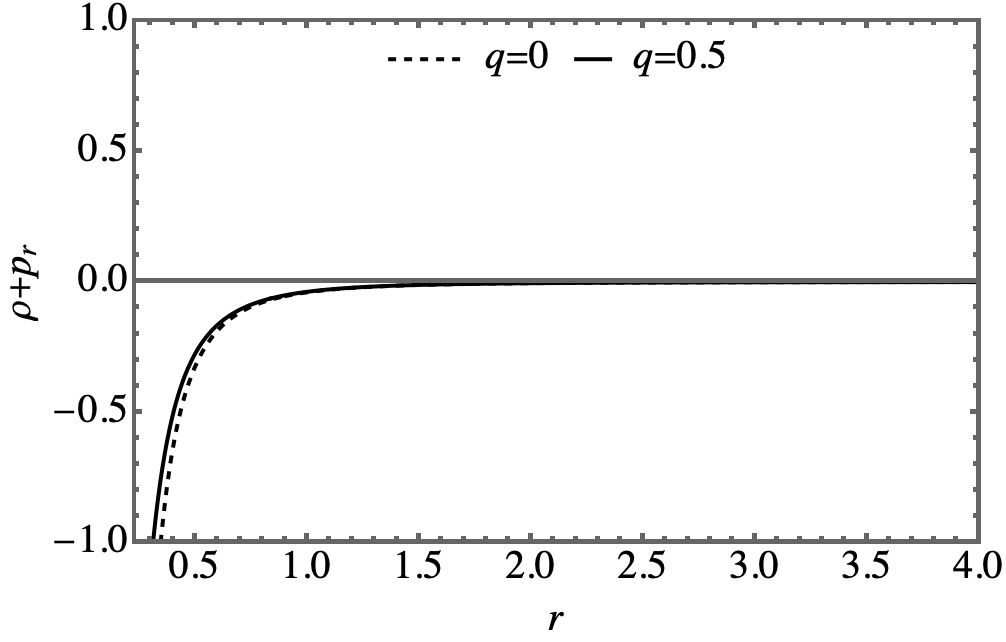}
    \includegraphics[width = 8.2 cm]{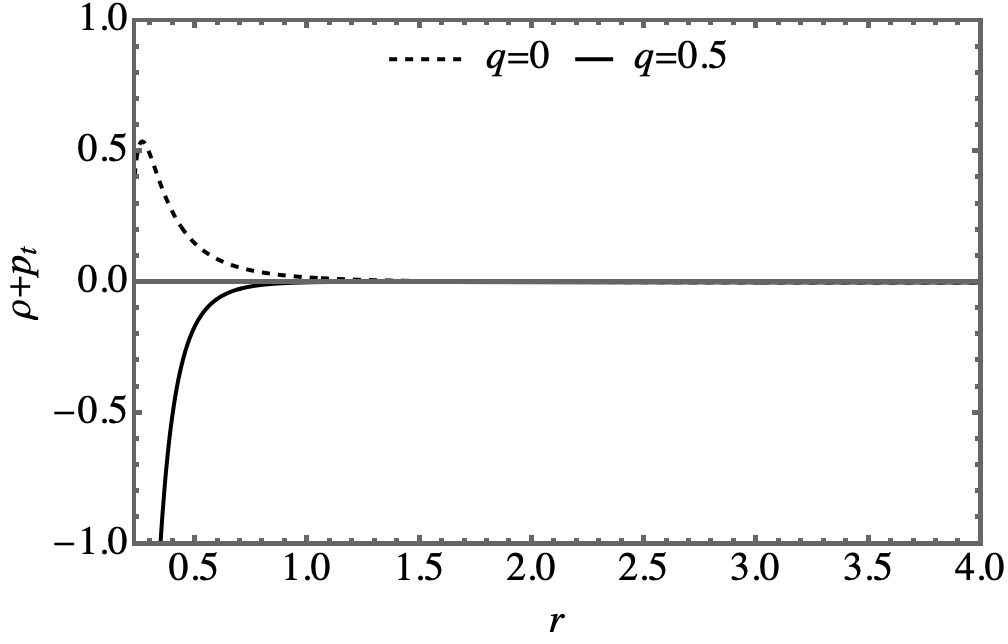}
    \includegraphics[width = 8.3 cm]{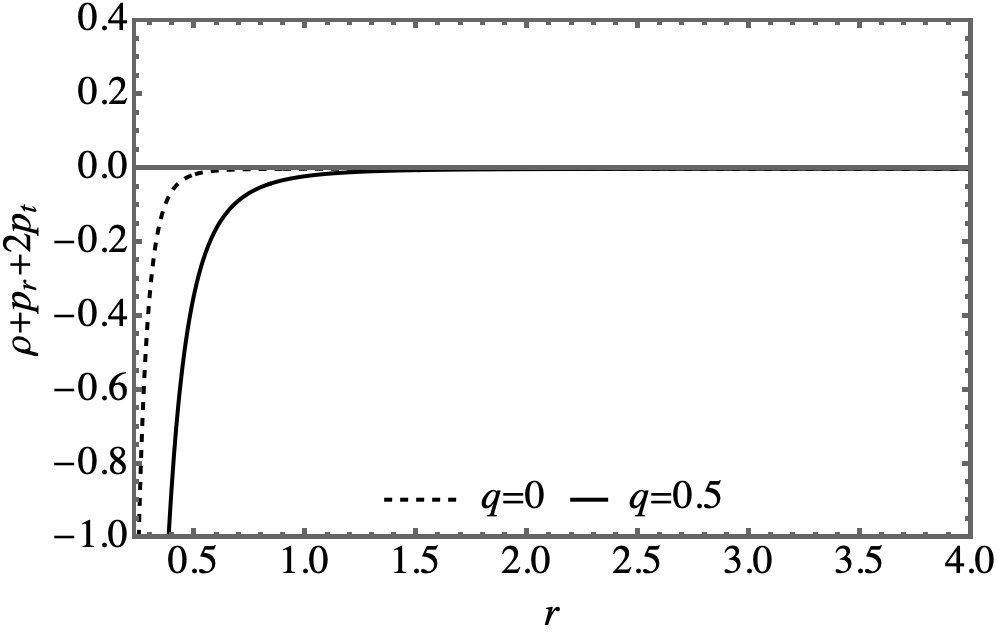}
    \caption{Figures show that the behavior of NEC and SEC diagrams have been plotted for the isotropic wormhole against $r$. Here also NEC is violated, this implies WEC is also violated. For plotting the constants are $\alpha = 0.001$ and $r_0 = 1$ along with $\omega = 0.5$.}
    \label{energyiso}
\end{figure}

\subsection{Anisotropic solution}
Using the shape function provided in Eq.(\ref{n26}), we can compute the energy-momentum components. Specifically, we determine the energy density and the radial component as follows:
\beq
\rho&=&\frac{1}{8 \pi  \alpha  B r^4}\Big(r_{0}^{-2 (\omega_{t} +1)} \big(\alpha  (2 (B-2) \omega_{t} +B-8) r^{2 \omega_{t} +2} \big(\alpha +q^2-r_{0}^2\big)\nonumber\\&&\quad\quad\quad\quad+3 (B-1) r^4 r_{0}^{2 \omega +2}+4 \alpha  q^2 r_{0}^{2 \omega_{t} +2}\big)\Big)\,,
\eeq
and for the tangential component
\beq
 p_{t}=\frac{\omega_{t}}{8 \pi }\Big(r^{2 (\omega_{t} -1)} r_{0}^{-2 (\omega_{t} +1)} \left(\alpha +q^2-r_{0}^2\right)\Big)\,,
\eeq
where
\beq
B\equiv \sqrt{1+4 \alpha  r^{2 (\omega_{t} -1)} r_{0}^{-2 (\omega_{t} +1)} \left(\alpha +q^2-r_{0}^2\right)-\frac{4 \alpha  q^2}{r^4}}.
\eeq
In the plots shown in Fig.(\ref{energyiso}), we observe that at the wormhole throat, $r =r_0$, the energy condition $(\rho+ p_{r})|_{r=r_{0}}< 0$ and the condition $(\rho+p_{r}+2p_{t})|_{r=r_{0}}< 0$ are satisfied by arbitrarily small values.
\begin{figure}[t]
    \centering
    \includegraphics[width = 8 cm]{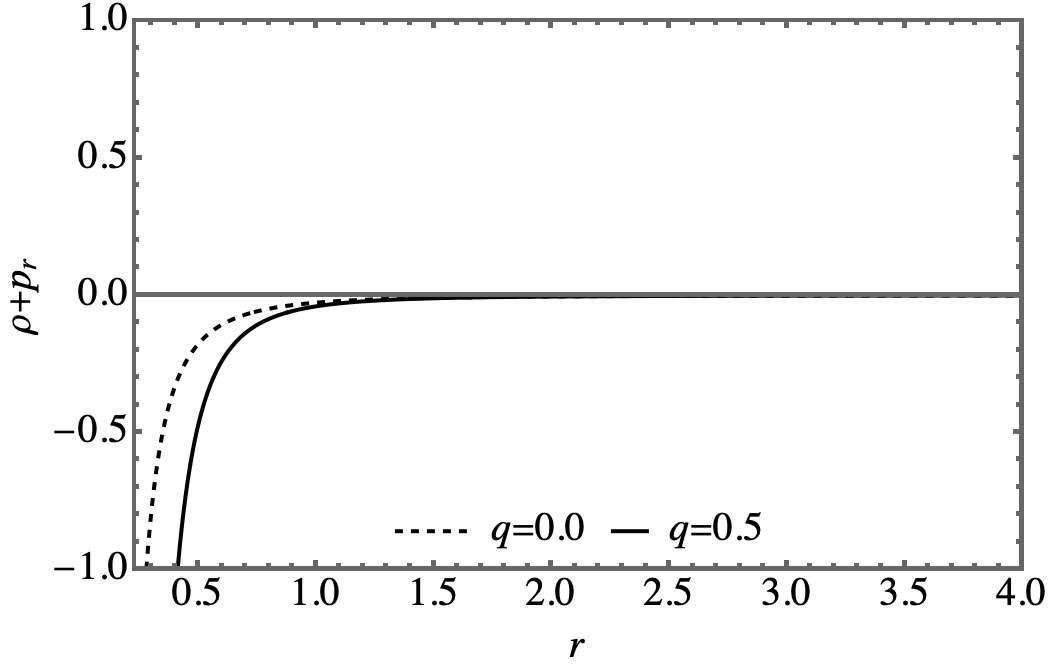}
    \includegraphics[width = 8.2 cm]{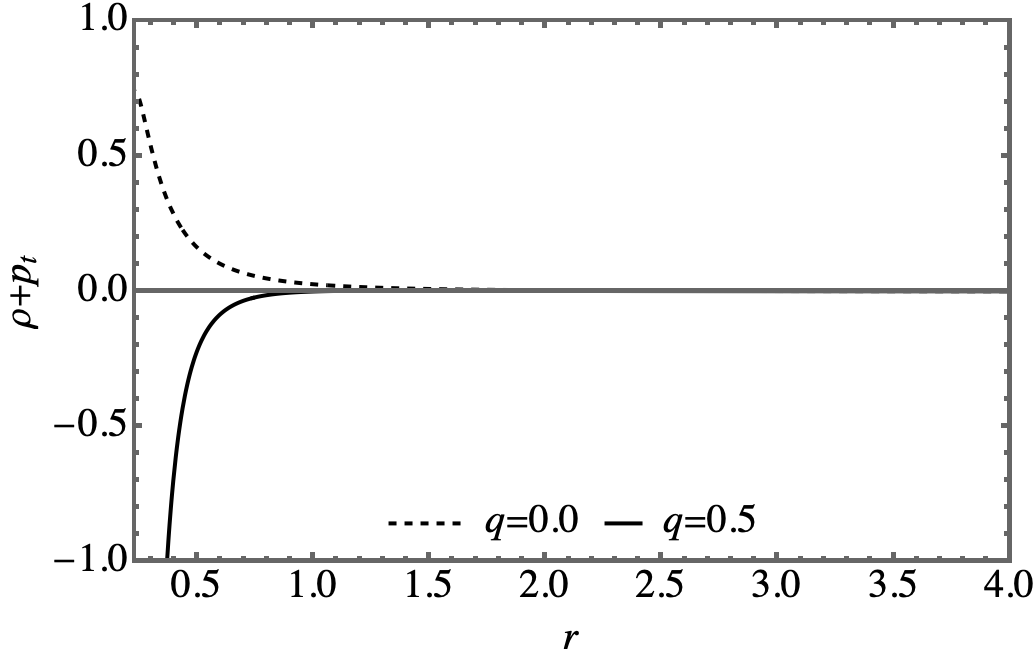}
    \includegraphics[width = 8.3 cm]{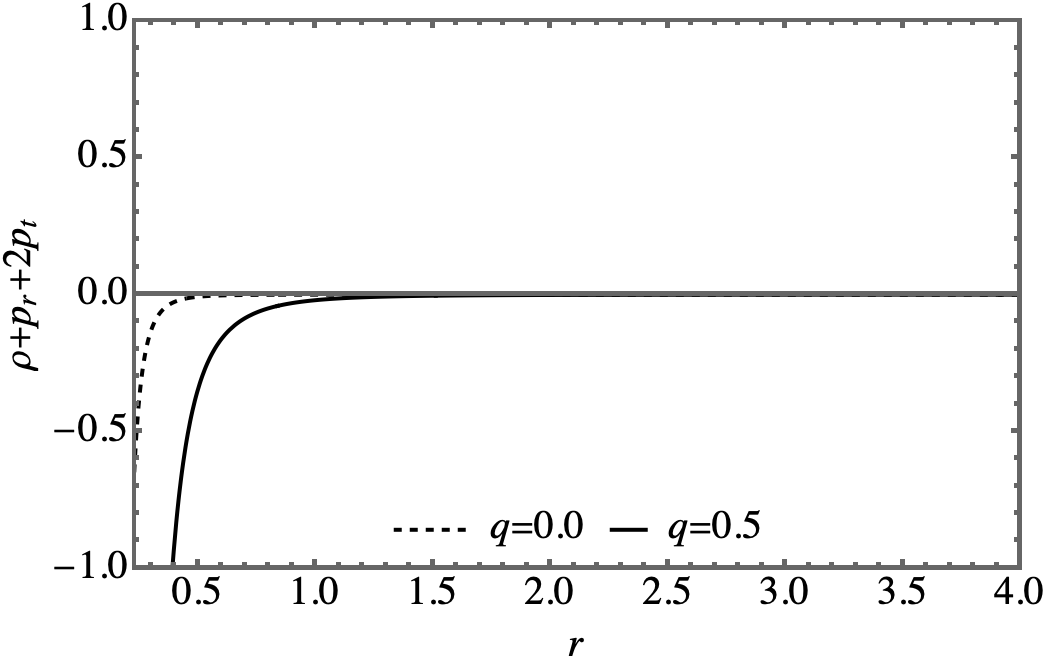}
    \caption{Figures show that the behavior of NEC and SEC diagrams have been plotted for the anisotropic wormhole against $r$. Here also NEC is violated, this implies WEC is also violated. For plotting the constants are $\alpha = 0.001$ and $r_0 = 1$ along with $\omega_t = -1/3$.}
    \label{energyiso}
\end{figure}

\subsection{Model with a specific energy density}
Using the shape function provided in Eq.(\ref{Anla}), we can compute the energy-momentum components. Specifically, we determine the energy density and the radial component as follows:
\beq
\rho&=&\rho_{0}\left(\frac{r_{0}}{r}\right)^{\beta }\,,\\
 p_{r}&=&\frac{1}{8 \pi }\Bigg(\frac{1}{\alpha }+\frac{\alpha +q^2+r_{0}^2}{r^3 r_{0}}-\frac{1}{\alpha}{\cal B}(q,\alpha,\rho_{0})+\frac{8 \pi\rho_{0}}{(\beta -3) r^3}\left(\rho_{0}^3-r^3 \left(\frac{\rho_{0}}{r}\right)^{\beta }\right)\Bigg)\,,
\eeq
and for the tangential component
\beq
 p_{t}&=&\frac{1}{16 \pi  \alpha  (\beta -3) B r^3 r_{0}}\Bigg(2 r^3 r_{0} \Bigg(-8 \pi  \alpha  \rho_{0} \left(\frac{r_{0}}{r}\right)^{\beta}\Bigg(\sqrt{\frac{{\cal C}(q,\alpha,\rho_{0})}{(\beta -3) r^4 r_{0}}}-4\Bigg)\nonumber\\&&+\beta  \Bigg(4 \pi  \alpha  \rho_{0} \left(\frac{r_{0}}{r}\right)^{\beta } \Bigg(\sqrt{\frac{{\cal C}(q,\alpha,\rho_{0})}{(\beta -3) r^4 r_{0}}}-2\Bigg)+\sqrt{\frac{{\cal C}(q,\alpha,\rho_{0})}{(\beta -3) r^4 r_{0}}}-1\Bigg)-3 \sqrt{\frac{{\cal C}(q,\alpha,\rho_{0})}{(\beta -3) r^4 r_{0}}}+3\Bigg)\nonumber\\&&-\alpha  \left((\beta -3) \left(\alpha +q^2+r_{0}^2\right)+8 \pi  \rho_{0} r_{0}^4\right) \left(\sqrt{\frac{{\cal C}(q,\alpha,\rho_{0})}{(\beta -3) r^4 r_{0}}}+2\right)\Bigg)\,,
\eeq
where we have defined new functions
\beq
{\cal B}(q,\alpha,\rho_{0})&\equiv& \sqrt{-\frac{4 \alpha  q^2}{r^4}+\frac{4 \alpha  \left(\alpha +q^2+\frac{8 \pi \rho_{0} r_{0}^4}{\beta -3}+r_{0}^2\right)}{r^3 r_{0}}-\frac{32 \pi  \alpha \rho_{0} \left(\frac{r_{0}}{r}\right)^{\beta }}{\beta -3}+1}\,,\\{\cal C}(q,\alpha,\rho_{0})&\equiv& 4 \alpha  r \left((\beta -3) \left(\alpha +q^2+r_{0}^2\right)+8 \pi  \rho_{0} r_{0}^4\right)-4 \alpha  (\beta -3) q^2 r_{0}\nonumber\\&&+r^4 r_{0} \left(\beta -32 \pi  \alpha  \rho_{0} \left(\frac{r_{0}}{r}\right)^{\beta }-3\right).
\eeq

\begin{figure}[t]
    \centering
    \includegraphics[width = 8 cm]{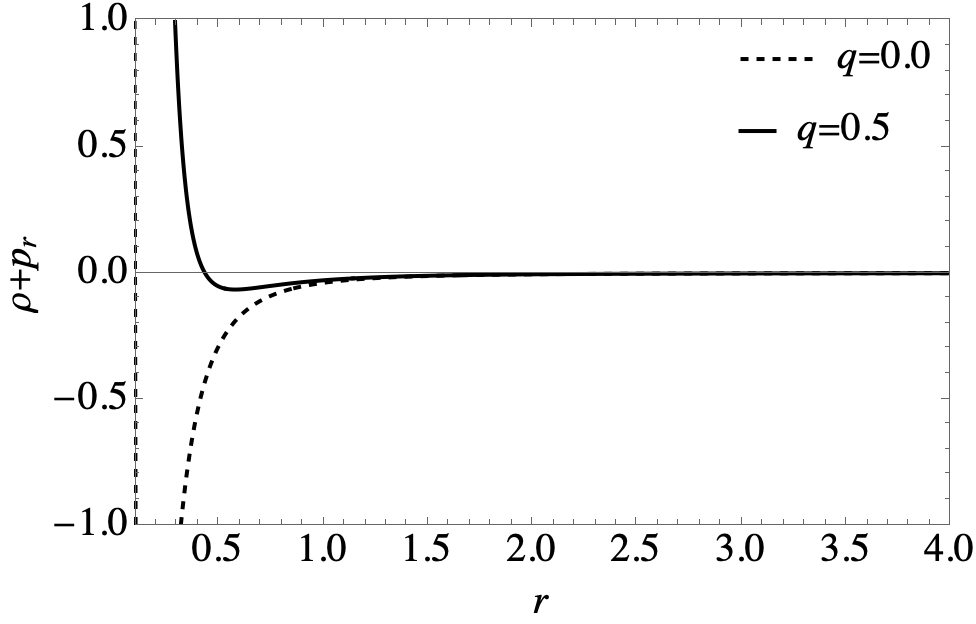}
    \includegraphics[width = 8.2 cm]{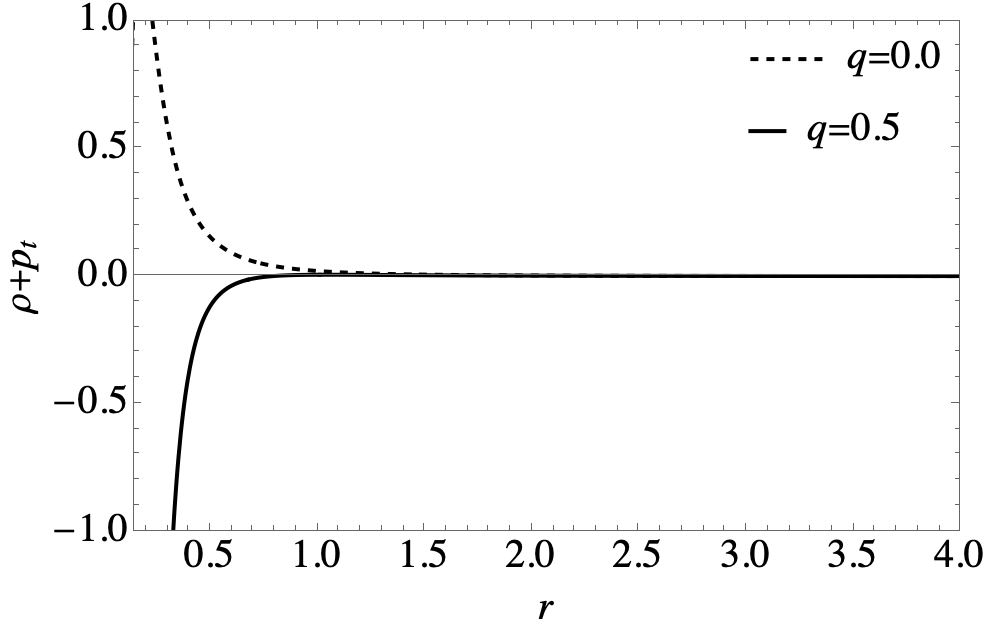}
    \includegraphics[width = 8.3 cm]{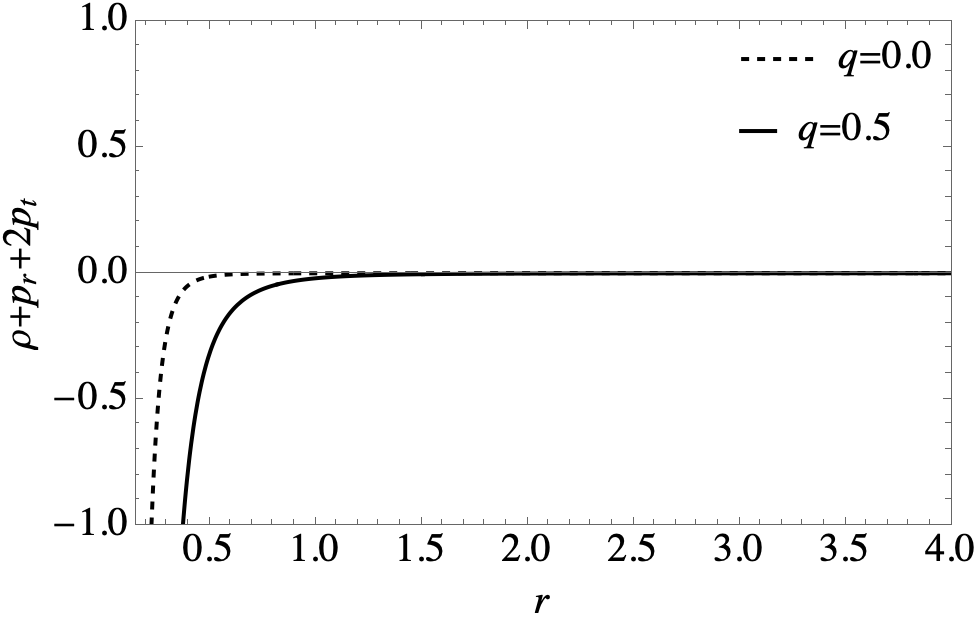}
    \caption{Figures show that the behavior of NEC and SEC diagrams have been plotted for the anisotropic wormhole against $r$. Here also NEC is violated, this implies WEC is also violated. For plotting the constants are $\alpha = 0.001$ and $r_0 = 1$ along with $\beta = 4$.}
    \label{energyiso1}
\end{figure}
In the plots shown in Fig.(\ref{energyiso1}), we also find that at the wormhole throat, $r =r_0$, the energy condition $(\rho+ p_{r})|_{r=r_{0}}< 0$ and the condition $(\rho+p_{r}+2p_{t})|_{r=r_{0}}< 0$ are satisfied by arbitrarily small values.

\section{Stability analysis}
\label{sec4}
The generalized Tolman–Oppenheimer–Volkov (TOV) equation extends the traditional TOV equation to include anisotropic pressures and exotic forms of matter. This can help examine aspects of wormhole stability, although it might not cover all unique features. By applying the generalized TOV equation to a wormhole, one can analyze equilibrium conditions under gravitational forces and pressure gradients from exotic matter. The equilibrium condition is given by the generalized TOV equation, expressed as follows:
\beq
\frac{dp_{r}}{dr}+\frac{1}{2}\big(\rho+p_{r}\big)\Phi'+\frac{2}{r}\big(p_{r}-p_{t}\big)=0\,.
\eeq
This helps in understanding the balance required to maintain a stable wormhole. For detailed study, we refer to Refs.\cite{Oppenheimer:1939ne,Tolman:1939jz,Sokoliuk:2021rtv,Dutta:2023wfg}. For our purpose, we recast the above equation as
\beq
F_{a}+F_{g}+F_{h}=0\,,\label{sumE}
\eeq
which provides the equilibrium condition for the wormholes. Here we have defined new quantities:
\beq
F_{h}=-\frac{dp_{r}}{dr}\,,\quad\quad F_{g}=-\frac{1}{2}\big(\rho+p_{r}\big)\Phi'\,,\quad\quad F_{a}=\frac{2}{r}\big(p_{t}-p_{r}\big)\,.
\eeq
Here, $F_a$ represents the force resulting from the anisotropic matter of the wormhole, $F_g$ denotes the gravitational force, and $F_h$ is the hydrostatic force. $F_a$ arises due to modifications in the gravitational Lagrangian of the Einstein-Hilbert action. However, as indicated by Eq.(\ref{sumE}), it is evident that for the system to remain in equilibrium, the sum of these three forces must equate to zero.

\begin{figure}[t]
    \centering
    \includegraphics[width = 8.1 cm]{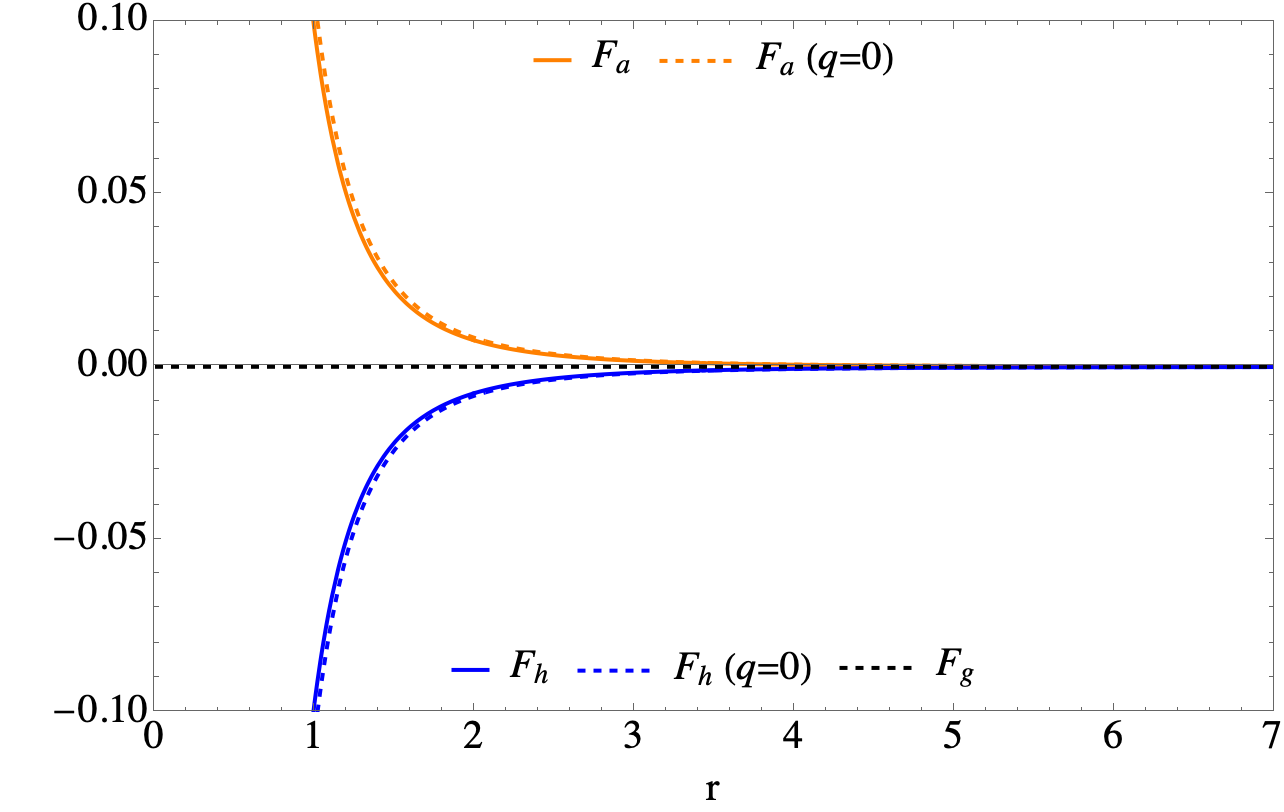}
    \includegraphics[width = 8.1 cm]{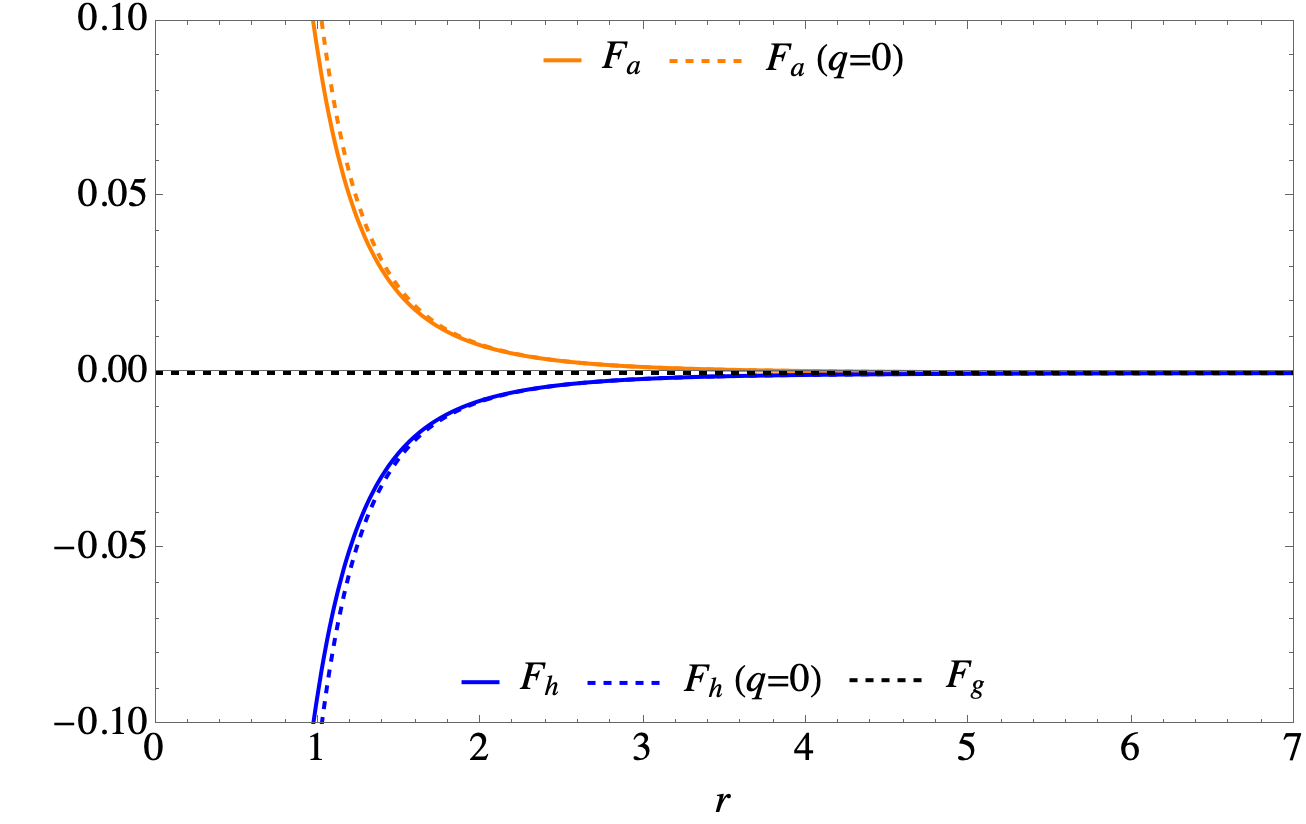}
    \caption{The three forces for the equilibrium conditions are plotted against $r$; Left panel: for $\omega_{t} = -1/3,\, \alpha = 0.001,\, r_{0} = 1,\, q = 0.3$, Right panel: for $\alpha = 0.001,\, \beta = 4,\, \rho_{0}= 0.01,\, q= 0.3,\, r_{0}= 1$. Here $F_{g}= 0$ for a constant redshift function. Note that the case $q=0$ is also shown for reference.}
    \label{sumE12}
\end{figure}

In the anisotropic solution, the behaviors of $F_a,\,F_{g}$, and $F_h$ are illustrated in Fig.(\ref{sumE12}) for specific parameter values where the energy conditions are met. Notably, $F_g$ is zero because $\Phi$ is assumed to be constant, meaning the gravitational force has no impact on our model. The figure shows that the other two forces are equal in magnitude but opposite to each other. This indicates that the equilibrium of forces is achieved through the combined effect of the three force terms, supporting the system's stability. The same behavior is observed in a model with a specific energy density.

\section{Conclusion}

In the present work, we constructed models of static wormholes within the framework of in 4-dimensional Einstein-Gauss-Bonnet (4D EGB) gravity an (an)isotropic energy momentum tensor (EMT) and a Maxwell field as supporting matters for the wormhole geometry assuming a constant redshift function. We obtained exact spherically symmetric wormhole solutions in 4D EGB gravity for an isotropic and anisotropic matter sources under the effect of electric charge. Furthermore, we examined the null, weak and strong energy conditions at the wormhole throat of radius $r=r_{0}$. We demonstrated that at the wormhole throat, the classical energy conditions are violated by arbitrary small amount. Additionally, we analyze the wormhole geometry incorporating semiclassical corrections through embedding diagrams.

From the viewpoint of the equilibrium condition for wormhole configuration, we considered the modified Tolman-Oppenheimer-Volkov equation. The equation can be devided into three parts: the anisotropic force $F_{a} = 2(p_{t}-p_{r})/r$, the hydrostatic force $F_{h}=-d p_{r}/dr$ and the force due to gravitational $F_{g}=-\big(\rho+p_{r}\big)\Phi'/2$, where the last one vanishes in our model since the redshift function is assumed to be constant. Taking the derivative of radial pressure, it is easy to check that $F_{h}= -F_{a}$ and hence we can deduce that the wormhole solutions are in equilibrium as the anisotropic and hydrostatic forces cancel each other

Furthermore, the results in this study generalize previous discussions in Ref.\cite{Jusufi:2020yus} by extending them beyond the limit $q\rightarrow 0$. The results also provide a broader generalization to Morris-Thorne wormholes of General Relativity, which are encompassed as a special case in the limits $\alpha\rightarrow 0$ and $q\rightarrow 0$. Exploring the possibility of extending wormholes to include rotation and applying them to more general Lovelock gravity theories [93] are current and intriguing research challenges actively under consideration. The observational aspects of wormhole solutions are also worthy of study, such as through the utilization of gravitational lensing effects, see e.g., Ref.\cite{Mehdizadeh:2018smu}, and  shadows and photon spheres in static and rotating traversable wormholes \cite{Tangphati:2023mpk} as well as a charge one \cite{Saleem:2024kld}.

\acknowledgments
This work is financially supported by Thailand NSRF via PMU-B under grant number PCB37G6600138.

\end{document}